\documentclass[sigplan,screen]{acmart}

\AtBeginDocument{%
  \providecommand\BibTeX{{%
    \normalfont B\kern-0.5em{\scshape i\kern-0.25em b}\kern-0.8em\TeX}}}

\copyrightyear{2024}
\acmYear{2024}
\setcopyright{acmlicensed}\acmConference[DAC '24]{61st ACM/IEEE Design Automation Conference}{June 23--27, 2024}{San Francisco, CA, USA}
\acmBooktitle{61st ACM/IEEE Design Automation Conference (DAC '24), June 23--27, 2024, San Francisco, CA, USA}
\acmDOI{10.1145/3649329.3658469}
\acmISBN{979-8-4007-0601-1/24/06}

\usepackage[ruled,vlined,linesnumbered]{algorithm2e}
\usepackage{listings}
\usepackage{xspace}
\usepackage{hyperref}
\usepackage{cleveref}

\usepackage{tikz}
\newcommand{\tikzxmark}{%
\tikz[scale=0.23] {
    \draw[red,line width=0.7,line cap=round] (0,0) to [bend left=6] (1,1);
    \draw[red,line width=0.7,line cap=round] (0.2,0.95) to [bend right=3] (0.8,0.05);
}}
\newcommand{\tikzcmark}{%
\tikz[scale=0.23] {
    \draw[blue,line width=0.7,line cap=round] (0.25,0) to [bend left=10] (1,1);
    \draw[blue,line width=0.8,line cap=round] (0,0.35) to [bend right=1] (0.23,0);
}}

\newcommand{\coolname}[1]{\textit{Specure}\xspace#1}%
\newcommand{\phaseone}[1]{\textsc{Offline Phase}\xspace#1}%
\newcommand{\phasetwo}[1]{\textsc{Online Phase}\xspace#1}%
\newcommand{\svs}[1]{speculative execution vulnerabilities\xspace#1}%
\newcommand{\sa}[1]{speculative execution attack\xspace#1}%

\newcommand{\vd}[0]{Vulnerability Detector\xspace}%
\newcommand{\hf}[0]{Hardware Fuzzer\xspace}%
\newcommand{\mv}[0]{Microarchitecture Visualizer\xspace}%
\newcommand{\coc}[0]{Coverage Calculator\xspace}%
\newcommand{\ld}[0]{Leakage Detector\xspace}%

\newcommand{\boom}{{\tt BOOM}}
\newcommand{\zenbleed}{{\textit{Zenbleed}}}

\newcommand\blfootnote[1]{%
  \begingroup
  \renewcommand\thefootnote{}\footnote{#1}%
  \addtocounter{footnote}{-1}%
  \endgroup
}

\begin{document}

\definecolor{codegreen}{rgb}{0,0.6,0}
\definecolor{codegray}{rgb}{0.5,0.5,0.5}
\definecolor{codepurple}{rgb}{0.58,0,0.82}
\definecolor{backcolour}{rgb}{0.95,0.95,0.92}

\let\othelstnumber=\thelstnumber
\def\createlinenumber#1#2{
    \edef\thelstnumber{%
        \unexpanded{%
            \ifnum#1=\value{lstnumber}\relax
              #2%
            \else}%
        \expandafter\unexpanded\expandafter{\thelstnumber\othelstnumber\fi}%
    }
    \ifx\othelstnumber=\relax\else
      \let\othelstnumber\relax
    \fi
}

\lstdefinestyle{customc}{
  belowcaptionskip=1\baselineskip,
  breaklines=true,
  frame=single,
  xleftmargin=0.35cm,
  xrightmargin=0.15cm,
  numbers=left,
  numbersep=5pt,  
  language=C,
  showstringspaces=false,
  basicstyle=\footnotesize\ttfamily,
  keywordstyle=\bfseries\color{green!40!black},
  commentstyle=\itshape\color{purple!40!black},
  identifierstyle=\color{blue},
  stringstyle=\color{orange},
}

\lstdefinestyle{customcArianeExploit1}{
  breaklines=true,
  frame=single,
  xleftmargin=0.4cm,
  xrightmargin=0.2cm,
  numbers=left,
  numbersep=5pt,  
  language=C,
  showstringspaces=false,
  basicstyle=\footnotesize\ttfamily,
  keywordstyle=\bfseries\color{green!40!black},
  commentstyle=\itshape\color{purple!60!black},
  identifierstyle=\color{blue},
  stringstyle=\color{yellow!50!black},
  morekeywords={asm},
  keywordstyle=[2]\bfseries\color{brown!60!black},
}

\lstdefinestyle{customasm}{
  belowcaptionskip=1\baselineskip,
  breaklines=true,
  frame=single,
  xleftmargin=0.4cm,
  xrightmargin=0.2cm,
  numbers=left,
  numbersep=5pt,
  xleftmargin=\parindent,
  language=[x86masm]Assembler,
  basicstyle=\footnotesize\ttfamily,
  keywordstyle=\bfseries\color{blue},
  commentstyle=\itshape\color{green!50!black},
  identifierstyle=\color{black},
  stringstyle=\color{brown},
}

\lstdefinestyle{customcArianeExploit}{
  breaklines=true,
  frame=single,
  xleftmargin=0.4cm,
  xrightmargin=0.2cm,
  numbers=left,
  numbersep=5pt,  
  language=C,
  showstringspaces=false,
  basicstyle=\footnotesize\ttfamily,
  keywordstyle=\bfseries\color{blue},
  commentstyle=\itshape\color{green!50!black},
  identifierstyle=\color{black},
  stringstyle=\color{brown},
  morekeywords={asm},
  keywordstyle=[2]\bfseries\color{black},
}

\lstdefinestyle{customlog}{
  breaklines=true,
  frame=single,
  xleftmargin=0.35cm,
  xrightmargin=0.15cm,
  numbers=left,
  numbersep=5pt,  
  language=C,
  showstringspaces=false,
  basicstyle=\footnotesize\ttfamily,
  keywordstyle=\color{blue},
  commentstyle=\itshape\color{purple!40!black},
  identifierstyle=\color{blue},
  stringstyle=\color{orange},
  keywords=[2]{INFO},
  keywords=[3]{ERROR},x
  keywordstyle=[2]\bfseries\color{green!40!black},
  keywordstyle=[3]\bfseries\color{red!500!black},
}

\definecolor{verilogcommentcolor}{RGB}{104,180,104}
\definecolor{verilogkeywordcolor}{RGB}{49,49,255}
\definecolor{verilogsystemcolor}{RGB}{128,0,255}
\definecolor{verilognumbercolor}{RGB}{255,143,102}
\definecolor{verilogstringcolor}{RGB}{160,160,160}
\definecolor{verilogdefinecolor}{RGB}{128,64,0}
\definecolor{verilogoperatorcolor}{RGB}{0,0,128}
\definecolor{pointcolor}{RGB}{192,0,0} %
\lstdefinestyle{prettyverilog}{
   language           = Verilog,
   commentstyle       = \color{verilogcommentcolor},
   alsoletter         = \$'0123456789\`,
   literate           = *{+}{{\verilogColorOperator{+}}}{1}%
                         {-}{{\verilogColorOperator{-}}}{1}%
                         {@}{{\verilogColorOperator{@}}}{1}%
                         {;}{{\verilogColorOperator{;}}}{1}%
                         {*}{{\verilogColorOperator{*}}}{1}%
                         {?}{{\verilogColorOperator{? }}}{1}%
                         {:}{{\verilogColorOperator{:}}}{1}%
                         {<}{{\verilogColorOperator{<}}}{1}%
                         {>}{{\verilogColorOperator{>}}}{1}%
                         {!}{{\verilogColorOperator{!}}}{1}%
                         {^}{{\verilogColorOperator{^}}}{1}%
                         {|}{{\verilogColorOperator{|}}}{1}%
                         {=}{{\verilogColorOperator{= }}}{1}%
                         {==}{{\verilogColorOperator{== }}}{1}%
                         {=>}{{\verilogColorOperator{=> }}}{1}%
                         {[}{{\verilogColorOperator{[}}}{1}%
                         {]}{{\verilogColorOperator{]}}}{1}%
                         {(}{{\verilogColorOperator{(}}}{1}%
                         {)}{{\verilogColorOperator{)}}}{1}%
                         {,}{{\verilogColorOperator{,}}}{1}%
                         {.}{{\verilogColorOperator{.}}}{1}%
                         {~}{{\verilogColorOperator{$\sim$}}}{1}%
                         {\%}{{\verilogColorOperator{\%}}}{1}%
                         {\&}{{\verilogColorOperator{\&}}}{1}%
                         {\&\&}{{\verilogColorOperator{\&\& }}}{1}%
                         {\#}{{\verilogColorOperator{\#}}}{1}%
                         {\ /\ }{{\verilogColorOperator{\ /\ }}}{3}%
                         {\ _}{\ \_}{2}%
                        ,
   morestring         = [s][\color{verilogstringcolor}]{"}{"},%
   identifierstyle    = \color{black},
   vlogdefinestyle    = \color{verilogdefinecolor},
   vlogconstantstyle  = \color{verilognumbercolor},
   vlogsystemstyle    = \color{verilogsystemcolor},
   basicstyle         = \scriptsize\fontencoding{T1}\ttfamily,
  columns=fullflexible, 
   keywordstyle       = \bfseries\color{verilogkeywordcolor},
   morekeywords      = {val, when, port, coverage, unique},
   numbers            = left,
   numbersep          = 5pt,
   tabsize            = 2,
   escapeinside       = {/*!}{!*/},
   upquote            = true,
   sensitive          = true,
   showstringspaces   = false, %
   frame              = single,
   breaklines         = true,
   abovecaptionskip   = 0pt,
   belowcaptionskip   = 2pt,   
   xleftmargin        =0.35cm,
   xrightmargin       =0.15cm,
   captionpos         = b,
   emph               = {Point, Point0, Point1, Point2, Point3, Point4, Point5, Point6, Point7, Point8, Point9},
   emphstyle          =\color{pointcolor},%
   emph               = {[2] STVEC,SCOUNTEREN,MSTATUS,MTVEC,ML1_ICACHE_MISS,ML1_DCACHE_MISS,MITLB_MISS,MDTLB_MISS,
                             MLOAD,MSTORE,MEXCEPTION,MEXCEPTION_RET,MBRANCH_JUMP,MCALL,MRET,MMIS_PREDICT,MSB_FULL,
                             MIF_EMPTY,MHPM_COUNTER_17,MHPM_COUNTER_18,MHPM_COUNTER_19,MHPM_COUNTER_20,MHPM_COUNTER_21,
                             MHPM_COUNTER_22,MHPM_COUNTER_23,MHPM_COUNTER_24,MHPM_COUNTER_25,MHPM_COUNTER_26,MHPM_COUNTER_27,
                             MHPM_COUNTER_28,MHPM_COUNTER_29,MHPM_COUNTER_30,MHPM_COUNTER_31}, %
   emphstyle          = {[2]\bfseries\color{verilogkeywordcolor}}
}

\makeatletter

\newcommand\language@verilog{Verilog}
\expandafter\lst@NormedDef\expandafter\languageNormedDefd@verilog%
  \expandafter{\language@verilog}
  
\lst@SaveOutputDef{`'}\quotesngl@verilog
\lst@SaveOutputDef{``}\backtick@verilog
\lst@SaveOutputDef{`\$}\dollar@verilog

\newcommand\getfirstchar@verilog{}
\newcommand\getfirstchar@@verilog{}
\newcommand\firstchar@verilog{}
\def\getfirstchar@verilog#1{\getfirstchar@@verilog#1\relax}
\def\getfirstchar@@verilog#1#2\relax{\def\firstchar@verilog{#1}}

\newcommand\addedToOutput@verilog{}
\lst@AddToHook{Output}{\addedToOutput@verilog}

\newcommand\constantstyle@verilog{}
\lst@Key{vlogconstantstyle}\relax%
   {\def\constantstyle@verilog{#1}}

\newcommand\definestyle@verilog{}
\lst@Key{vlogdefinestyle}\relax%
   {\def\definestyle@verilog{#1}}

\newcommand\systemstyle@verilog{}
\lst@Key{vlogsystemstyle}\relax%
   {\def\systemstyle@verilog{#1}}

\newcount\currentchar@verilog

\newcommand\@ddedToOutput@verilog
{%
   \ifnum\lst@mode=\lst@Pmode%
      \expandafter\getfirstchar@verilog\expandafter{\the\lst@token}%
      \expandafter\ifx\firstchar@verilog\backtick@verilog
         \let\lst@thestyle\definestyle@verilog%
      \else
         \expandafter\ifx\firstchar@verilog\dollar@verilog
            \let\lst@thestyle\systemstyle@verilog%
         \else
            \expandafter\ifx\firstchar@verilog\quotesngl@verilog
               \let\lst@thestyle\constantstyle@verilog%
            \else
               \currentchar@verilog=48
               \loop
                  \expandafter\ifnum%
                  \expandafter`\firstchar@verilog=\currentchar@verilog%
                     \let\lst@thestyle\constantstyle@verilog%
                     \let\iterate\relax%
                  \fi
                  \advance\currentchar@verilog by \@ne%
                  \unless\ifnum\currentchar@verilog>57%
               \repeat%
            \fi
         \fi
      \fi
   \fi
}

\lst@AddToHook{PreInit}{%
  \ifx\lst@language\languageNormedDefd@verilog%
    \let\addedToOutput@verilog\@ddedToOutput@verilog%
  \fi
}

\newcommand{\verilogColorOperator}[1]
{%
  \ifnum\lst@mode=\lst@Pmode\relax%
   {\bfseries\textcolor{verilogoperatorcolor}{#1}}%
  \else
    #1%
  \fi
}

\makeatother

\lstdefinestyle{mystyle}{
    commentstyle=\textit,
    keywordstyle=\textbf,
    stringstyle=\color{codepurple},
    basicstyle=\ttfamily,
    breakatwhitespace=false,         
    breaklines=true,      
    frame=single, 
    framexleftmargin=\parindent,
    captionpos=b,                    
    keepspaces=true,                 
    numbers=left,    
    numberstyle=\normalsize,
    stepnumber=1,
    numbersep=5pt,   
    xleftmargin=1.5\parindent,
    showspaces=false,                
    showstringspaces=false,
    showtabs=false,                  
    tabsize=2
}

\lstset{style=mystyle}

\lstset{
  language=Java, 
  basicstyle=\small, 
  frame=single, 
  breaklines=true, 
  postbreak=\raisebox{0ex}[0ex][0ex]{\ensuremath{\hookrightarrow\space}},
  deletestring=[b]",
  deletestring=[b]'
}

\lstdefinelanguage[RISC-V]{Assembler}
{
  alsoletter={.}, %
  alsodigit={0x}, %
  morekeywords=[1]{ %
    lb, lh, lw, lbu, lhu,
    sb, sh, sw,
    sll, slli, srl, srli, sra, srai,
    add, addi, sub, lui, auipc,
    xor, xori, or, ori, and, andi,
    slt, slti, sltu, sltiu,
    beq, bne, blt, bge, bltu, bgeu,
    j, jr, jal, jalr, ret,
    scall, break, nop
  },
  morekeywords=[2]{ %
    .align, .ascii, .asciiz, .byte, .data, .double, .extern,
    .float, .globl, .half, .kdata, .ktext, .set, .space, .text, .word
  },
  morekeywords=[3]{ %
    zero, ra, sp, gp, tp, s0, fp,
    t0, t1, t2, t3, t4, t5, t6,
    s1, s2, s3, s4, s5, s6, s7, s8, s9, s10, s11,
    a0, a1, a2, a3, a4, a5, a6, a7,
    ft0, ft1, ft2, ft3, ft4, ft5, ft6, ft7,
    fs0, fs1, fs2, fs3, fs4, fs5, fs6, fs7, fs8, fs9, fs10, fs11,
    fa0, fa1, fa2, fa3, fa4, fa5, fa6, fa7
  },
  morecomment=[l]{;},   %
  morecomment=[l]{\#},  %
  morestring=[b]",      %
  morestring=[b]'       %
}

\definecolor{mauve}{rgb}{0.58,0,0.82}

\lstdefinestyle{myriscv}{
  literate={ö}{{\"o}}1
           {ä}{{\"a}}1
           {ü}{{\"u}}1,
  basicstyle=\tiny\ttfamily,                    %
  breaklines=true,                              %
  commentstyle=\itshape\color{green!50!black},  %
  keywordstyle=[1]\color{blue!80!black},        %
  keywordstyle=[2]\color{orange!80!black},      %
  keywordstyle=[3]\color{red!50!black},         %
  stringstyle=\color{mauve},                    %
  identifierstyle=\color{teal},                 %
  frame=l,                                      %
  language=[RISC-V]Assembler,                   %
  tabsize=4,                                    %
  showstringspaces=false                        %
}

\title[Hybrid Speculative Vulnerability Detection]{Lost and Found in Speculation:
\\Hybrid Speculative Vulnerability Detection}

\author{Mohamadreza Rostami$^\dagger$, Shaza Zeitouni$^\dagger$, Rahul Kande$^\ddagger$, Chen Chen$^\ddagger$, Pouya Mahmoody$^\dagger$, Jeyavijayan (JV) Rajendran$^\ddagger$, Ahmad-Reza Sadeghi$^\dagger$}

\affiliation{$^\dagger$Technical University of Darmstadt, Darmstadt, Hessen, \country{Germany}\>\> $^\ddagger$Texas A\&M University, College Station, Texas, USA}

\renewcommand{\shortauthors}{M.Rostami and et.al.}

\begin{abstract}

Microarchitectural attacks represent a challenging and persistent threat to modern processors, exploiting inherent design vulnerabilities in processors to leak sensitive information or compromise systems. Of particular concern is the susceptibility of Speculative Execution, a fundamental part of performance enhancement, to such attacks.

We introduce \coolname, a novel pre-silicon verification method composing hardware fuzzing with Information Flow Tracking (IFT) to address speculative execution leakages. Integrating IFT enables two significant and non-trivial enhancements over the existing fuzzing approaches: i) automatic detection of microarchitectural information leakages vulnerabilities without golden model and ii) a novel \textit{Leakage Path} coverage metric for efficient vulnerability detection. 
\coolname identifies previously overlooked speculative execution vulnerabilities on the RISC-V \boom{} processor and explores the vulnerability search space 6.45$\times$ faster than existing fuzzing techniques. Moreover, \coolname detected known vulnerabilities 20$\times$ faster.

\end{abstract}



\maketitle

\section{Introduction}
\label{sec:intro}
\blfootnote{Email: $^\dagger$\{mohamadreza.rostami,shaza.zeitouni,pouya.mahmoody,ahmad.sad- eghi\}@trust.tu-darmstadt.de\>\>$^\ddagger$\{rahulkande,chenc,jv.rajendran\}@tamu.edu}
Hardware is the foundation of computing systems, and as such, insecure hardware exposes all critical systems to risk.
In particular, attacks based on microarchitectural design and implementation flaws in processors constitute a severe threat in exploiting the hardware to extract sensitive data or compromise the whole computing system
\cite{spectre,wait_for_it,aepic,cache_attack_tlb,hardfails,zenbleed,moghimi2023downfall}.
As designs advance and attackers evolve, we encounter increasingly sophisticated microarchitectural attacks~\cite{zenbleed, moghimi2023downfall}.

A crucial form of microarchitectural attacks exploits \textit{Speculative Execution}, a fundamental part of processors' performance enhancement~\cite{wait_for_it, aepic,moghimi2023downfall,zenbleed}.
Speculative execution technique predicts instruction outcomes before completion, reducing memory latency and mitigating pipeline stalls. However, speculative execution can lead to incorrect path forecasts, necessitating a rollback of changes made by transient instructions. 
Despite the rollback, some effects will still persist, serving as the root cause of speculative vulnerabilities. These vulnerabilities have been exploited to compromise systems' security by leveraging different channels to leak information from the microarchitectural layer to the architectural layer. These channels include direct channels such as architectural registers (e.g., Advanced Vector Extensions (AVX) registers \cite{zenbleed}, \texttt{EFLAGS} register \cite{wait_for_it}, and Memory-Mapped-Input/Output (MMIO) interfaces \cite{aepic}), as well as timing side/covert channels \cite{moghimi2023downfall,cache_attack_tlb}.
Researchers and manufacturers are investigating new defenses and countermeasures to protect against these attacks.
However, unlike software vulnerabilities that can be patched digitally, hardware patches often require physical modifications or replacements. Thus, detecting such vulnerabilities before fabrication is imperative.

However, existing approaches for automated detection of speculative execution vulnerabilities suffer from various deficiencies: They (i) are often restricted to known attack vectors \cite{introspectre, specdoctor, hide_and_seek_with_spectres}, hence, missing opportunities for detecting unknown vulnerabilities and their underlying causes, (ii) focus on specific microarchitectural elements rather than covering the entire processor-under-test~(PUT) \cite{hide_and_seek_with_spectres}, (iii) suffer from state explosion \cite{fadiheh2022exhaustive}, (iv) require additional hardware components \cite{specdoctor}, or (v) require a golden reference model~\cite{thehuzz}. These limitations reduce flexibility and increase manual effort and hardware overhead. Nonetheless, a more comprehensive and scalable technique for detecting hardware vulnerabilities is crucial to meet the field's evolving needs.

\textbf{Our Contributions.} We present a novel pre-silicon verification method, \coolname, effectively addressing critical aspects of speculative execution vulnerability detection. 
\coolname, a hardware-agnostic and non-invasive solution, detects speculative execution vulnerabilities in real-world size processors without necessitating additional hardware, modifications, or the use of golden reference models.
By composing hardware fuzzing and Information Flow Tracking (IFT), our approach efficiently identifies and quantifies direct channels within the processor design, revealing potential information leakage pathways and pinpointing their root cause in the hardware design. Overcoming challenges in coherent integration, we develop a strategy to express these channels and establish a novel coverage feedback mechanism for the fuzzer. 
Introduced coverage metric differs significantly from previous methods that solely desired to increase code coverage, blindly relying on luck to trigger vulnerabilities, thereby ensuring a comprehensive exploration of processor design and potential vulnerabilities. 
Our key contributions include:

\noindent\textbf{Automated detection of direct channels} and their root causes: \coolname leverages IFT to visualize the direct channel leakages between the processor's microarchitectural and architectural components. This is done by deriving the processor's Information Flow Graph (IFG) and extracting pathways from microarchitectural to architectural registers.

\noindent\textbf{Novel coverage metric.} We propose a unique \textit{Leakage Path} (LP) coverage for speculative execution leakages.
LP coverage provides two advantages over traditional code coverage-based fuzzers: (i) it identifies new and unknown vulnerabilities due to its fine-grained IFT-based guidance, and (ii) it discovers vulnerabilities more efficiently than other fuzzers.

\noindent\textbf{Implementation.} We implement \coolname using both open-source and commercial RTL simulators. To demonstrate its effectiveness, we use \boom{}~\cite{boom}, the most advanced open-source out-of-order processor, as the PUT. 
    
\noindent\textbf{Evaluation.} 
We validate our approach on \boom{} by coverage analysis and vulnerability detection.
With the novel coverage metric, \coolname explores the vulnerability search space $6.45 \times$ and detects known vulnerabilities $20 \times$ faster than existing fuzzing techniques.
We further validate our approach to detect recently discovered vulnerabilities \cite{wait_for_it,zenbleed} that remain undetectable by current methods. For that, we successfully reproduced the speculative execution vulnerabilities \cite{wait_for_it,zenbleed} in x86 architecture on \boom{}.

As we will explain in detail in the evaluation section (\autoref{sec:evaluation}), recent critical speculative execution vulnerabilities on commercial processors are derived from advanced microarchitectural features \cite{moghimi2023downfall,spectre,zenbleed,wait_for_it,aepic}. It is essential to stress that due to the absence of these features in open-source RISC-V processors, we emulate the behaviors of two recent and critical vulnerabilities on \boom{}. This demonstration aims to highlight the effectiveness of our hybrid fuzzer and the novel coverage metric in discovering and analyzing the root causes of such vulnerabilities.

\section{Background \& Related Work} 
\label{sec:background}

\textbf{Direct Channels for Speculative Execution Attacks.} 
Direct channels enable attackers to leak microarchitectural information directly to the architecture level without relying on timing or power side channels, which require precise measurements and side-channel interfaces \cite{aepic,wait_for_it, zenbleed, moghimi2023downfall}. %
Zhang et al. \cite{wait_for_it} showcased the exploitation of vulnerable \texttt{umonitor} and \texttt{umwait} instructions on recent Intel processors. These instructions optimize power states without interrupts, allowing information leakage during transient execution. \zenbleed{}~\cite{zenbleed} exploits zeroing registers optimization in AMD CPUs. When calling the zeroing instruction \texttt{vzeroupper} in the mispredicted speculative window, it will zero the upper bits of the register vector and deallocate dependencies of the vector and can allocated by the victims thread. \zenbleed{} was detected by fuzzing, showing the effectiveness of fuzzing in detecting vulnerabilities in large-scale designs.

\textbf{Fuzzing} or fuzz testing provisions test inputs to the software/hardware under test to uncover faults or vulnerabilities that traditional testing methods may miss \cite{afl++}. The fuzzer generates random, malformed, or unusual inputs to test how the program handles them. The initial set of test inputs, \textit{seeds}, can be automatically generated or manually crafted by verification engineers. 
During each fuzzing round, the fuzzer mutates the optimal test inputs from the preceding round using operations like \textit{bit/byte flipping}, \textit{swapping}, \textit{deleting}, or \textit{cloning} to generate new inputs.
Fuzzing has gained prominence in the security community due to its automation, cost-effectiveness, and scalability \cite{rfuzz,thehuzz,specdoctor,hide_and_seek_with_spectres}. 

\textbf{Fuzzing for Speculative Vulnerabilities.} %
Specdoctor~\cite{specdoctor} introduced a pre-silicon fuzzer that utilizes differential fuzzing with varied secret values to detect speculative execution vulnerabilities. It identifies potentially vulnerable modules based on known attacks and monitors them for discrepancies during fuzzing. However, Specdoctor's restricted selection criteria of known potentially vulnerable modules limits its ability to confirm exploitability and overlooks vulnerabilities.
On the other hand, Introspectre's \cite{introspectre} approach to detecting Meltdown-like vulnerabilities demands significant manual effort for preparing Meltdown gadgets, i.e., code snippets. The fuzzer is only responsible for randomly selecting gadgets, assigning random values to parameters within gadgets, and connecting gadgets to make a test case. Note that both works \cite{specdoctor,introspectre} have been evaluated using \boom{}.
In \cite{hide_and_seek_with_spectres}, a post-silicon fuzzer to detect speculative vulnerabilities in x86 CPUs is presented. To identify vulnerabilities, \cite{hide_and_seek_with_spectres} determines whether the fuzzer can induce identical architectural states with differing data cache states. This work is limited to detecting Spectre-like vulnerabilities and capturing timing leakage only in the L1 Data cache.

\textbf{Information Flow Tracking (IFT)} ensures sensitive data is not leaked or improperly accessed by monitoring the flow of information within a system. IFT tags data with labels that indicate their security level and tracks these tags as data moves through different components. Based on security specifications, IFT can inform if sensitive data leaks to insecure destinations~\cite{ift_survay}. Solt et al. \cite{ solt_cellift_2022} presented a dynamic IFT tool for detecting Spectre and Meltdown attacks. However, this approach has runtime overhead and requires changes to the underlying processor.

\begin{figure*}
    \centering
    \includegraphics[width=0.72\linewidth]{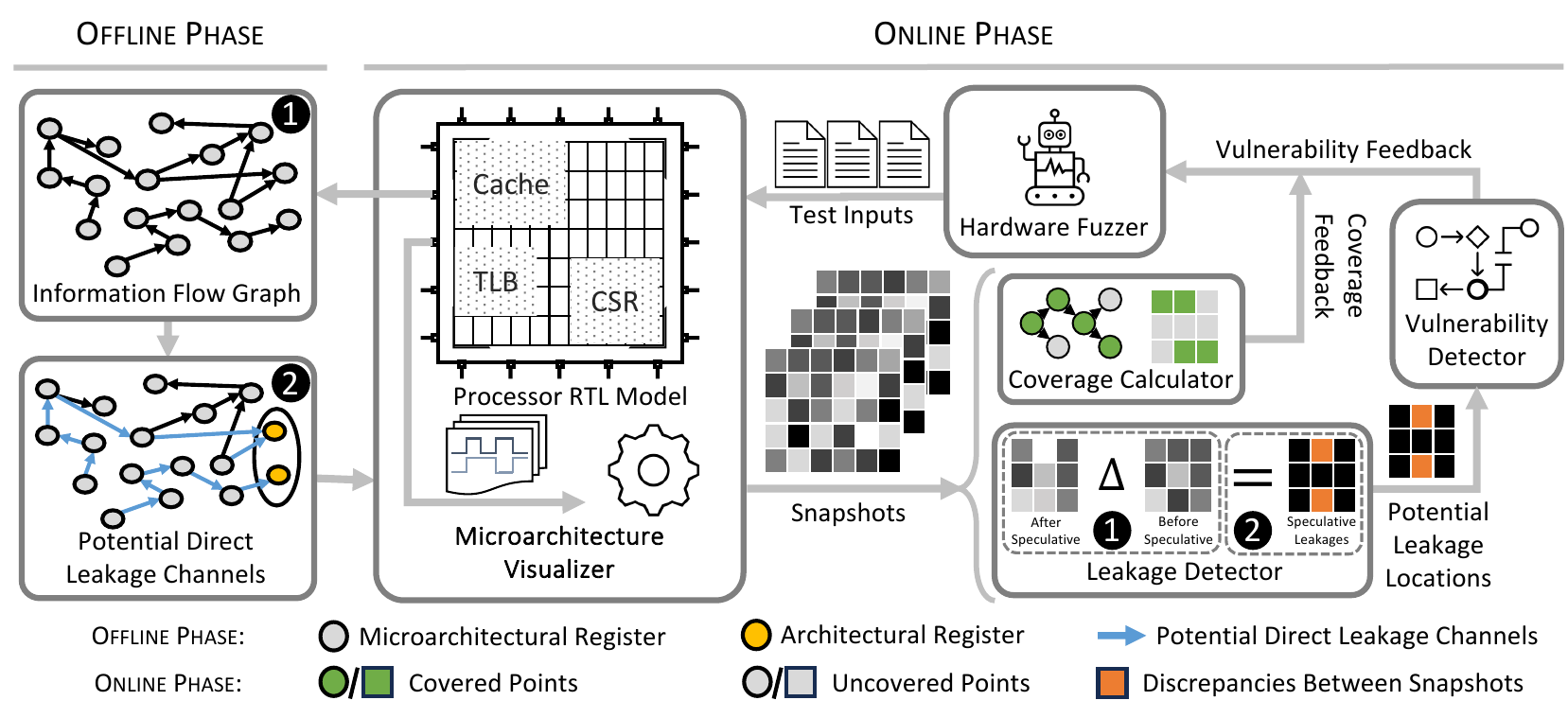}
    \caption{Overview of \coolname. The \phaseone leverages IFT technique to generate the \textit{IFG} and \textit{PDLC} of an RTL model. The \phasetwo leverages them to identify the existence and pinpoint the locations of speculative vulnerabilities.}
    \label{fig:framework}
    \vspace{-0.2cm}
\end{figure*}

\section{Design}\label{sec:design}
This section introduces our hybrid approach that combines fuzzing and IFT to identify \svs in the pre-silicon. We focus in this work on the detection of direct leakage channels.
A \sa requires first leaking information to microarchitectural components within the speculative window and then retrieval of leaked information from the microarchitectural layer to the architectural layer. 
To detect vulnerabilities triggered by speculative execution, \coolname entails two phases.
In the \phaseone (\autoref{sec:phaseone}), \coolname employs static IFT to identify potential leakage locations and paths within the PUT that could directly leak information from microarchitectural registers to architectural registers. 
In the \phasetwo (\autoref{sec:phasetwo}), \coolname employs fuzzing to generate inputs that trigger speculative execution. Then, \coolname detects leakage points within the PUT using a hyper-property aligned with the \svs previously introduced definition. 

\subsection{\phaseone~}%
\label{sec:phaseone}

This phase leverages IFT to extract information flows within the PUT. The process is performed statically, using the processor's RTL model, and in two steps, as shown in~\autoref{fig:framework}.

In \textbf{Step 1}, the information flow graph (IFG) is extracted from the PUT's RTL model. IFG is a directed graph that accurately represents the information flow within the hardware RTL model expressed as $IFG = (R, F)$. \({R}\) represents the set of all signals in the PUT. \(F\) represents the connections between signals defined as $F \subseteq \{(v,u)\ |\ v, u\  \in \ R\ and\ v \neq u\}$, where $(v,u)$ represents an edge or a connection between vertex $v$, the source signal, and vertex $u$, the destination.
\begin{lstlisting}[caption={Example RTL code: a \textit{top} module with two D-FFs.}, label=lst:exampleIFG, style={prettyverilog},captionpos=t, float=h,belowskip=-10pt] 
module D_FF(input d, input clk, output q);
 	reg q;
	always @(posedge clk) 
		q <= d;
endmodule
module top(input clk, input i, output o);
  reg q1;  
  D_FF df1(.d(i), .clk(clk), .q(q1));
  D_FF df2(.d(q1), .clk(clk), .q(o));
endmodule
\end{lstlisting}

\autoref{lst:exampleIFG} shows a simple RTL example, a \textit{top} module that implements a two clock-cycle delay using two D-flip-flops (D-FFs). %
In the $IFG$, \({R}\) is defined as:
\begin{align*}
    R = \{&top.q1,\ top.clk,\ top.i,\ top.o,\ top.df1.d,\ top.df1.q,\ \\
    &top.df1.clk,\ top.df2.d,\ top.df2.clk,\ top.df2.q\},
\end{align*}
and \({F}\) is defined as:
\begin{align*}
    F = \{&(top.clk,\ top.df1.clk), (top.clk,\ top.df2.clk),\\
    &(top.i,\ top.df1.d), (top.df1.d,\ top.df1.q),\\
    &(top.df1.q,\ top.q1), (top.q1,\ top.df2.d),\\
    &(top.df2.d,\ top.df2.q),(top.df2.q,\ top.o)\}.
\end{align*}

In \textbf{Step 2}, the PUT's IFG is used to extract all potential direct leakage channels (\textit{PDLC}) by extracting all paths from every microarchitectural register to all architectural registers. A direct leakage channel can be visualized in the IFG as a chain of edges starting from a microarchitectural register and ending in an architectural register. 
\coolname first identifies and labels all architectural registers in the set of all signals \({R}\) obtained in Step 1.
To distinguish the architectural registers from the microarchitectural registers in \({R}\), we parsed the latest privileged and unprivileged RISC-V instruction set architecture~(ISA) specification \cite{risc_v_isa_0} and automatically extracted programmer-accessible registers, including memory-mapped registers.
We apply the skewed-aware join approach 
to reduce the computation complexity of path extraction.
We reverse all paths in the IFG and search for routes that start from architectural registers to all microarchitectural registers,
This approach reduces the algorithm's complexity from \( O(V^2) \) to \( O(V) \), where \(V\) is the number of all registers in the PUT (i.e., \({|R|}\)).
 The output of \phaseone includes the PUT's IFG and the \textit{PDLC} list.

\subsection{\phasetwo~}%
\label{sec:phasetwo}

The processing steps of \coolname's \phasetwo consist of five components as shown in~\autoref{fig:framework}. 
In the following, we provide a detailed description of each component.

\textbf{\hf} aims to effectively explore the search space for potential vulnerabilities and cover as many items in the PDLC list as possible. In the scope of this work, \textit{search space} pertains to the collection of all possible combinations of instructions that cause the PUT to enter a state where it leaks information during the speculative execution window.
To boost vulnerability detection, we integrated special input seeds into the fuzzer alongside random seeds. The special seeds have transient execution windows covering scenarios like branch misprediction, branch target injection, and return stack buffer manipulation. %
Incorporating such seeds into the initial list accelerates the discovery of transient execution leaks, thus, enabling the fuzzer to faster trigger vulnerabilities compared to relying solely on random seeds.

\textbf{\mv} receives test inputs from the \hf and simulates them on the PUT. 
Then, it extracts the typical code coverage metrics (toggle, branch, finite-state machine~(FSM), etc.), execution traces, and waveforms that show PUT's signal values for each simulation clock cycle.
Using the waveforms, snapshots are generated from the microarchitectural and architectural signals. 
Each snapshot corresponds to a single clock cycle and contains the values of the PUT's signals, i.e., state, at that clock cycle. The snapshot represents the PUT's states. %

\textbf{\ld} identifies potential information leakage locations within speculative windows in two steps: 

In \textbf{Step 1}, the start and end of each speculative window are defined. This can be done by tracing speculative execution indicators, such as the processor's Re-order Buffer (RoB), which ensures the in-order commitment of executed instructions in an out-of-order processor.
For example, the RoB's in-queue of \boom{} receives multiple micro-operations. Each, equivalent to an instruction in \boom{}, contains a signal, \texttt{unsafe}, indicating whether this micro-operation starts a speculative window.
To find the end cycle of that speculative window, RoB receives signals, such as \texttt{brupdate}, from the branch predictor to confirm the (mis)prediction. 
Using these signals, which can be easily extracted from all snapshots generated by the \mv, we could detect each speculative window's start and end clock cycles.
We use this information and maintain a table, called \textit{Misspeculation Table} (\textit{MST}), that keeps the start and end clock cycles and the related instruction for each misspeculated window, as illustrated in \autoref{tab:specwindow}. 
Inputs generated by the \hf could have zero or more speculative windows.

\begin{table}[t]
\caption{The start and end cycles of Misspeculated Windows.}
\label{tab:specwindow}
    \resizebox{\columnwidth}{!}{%
    \begin{tabular}{c|c|c|c|l}
    \textbf{ID} & \textbf{Start} & \textbf{End} & \textbf{Instruction} & \multicolumn{1}{c}{\textbf{Instruction(Readable)}} \\ \hline
    \texttt{1}   & \texttt{34594}     & \texttt{34625}     & \texttt{FBEC52E3}      & \texttt{BGE S8, T5, 0x800025B0}\\
    \texttt{2}   & \texttt{89991}     & \texttt{90121}     & \texttt{FB6F42E3}      & \texttt{BLT T5, S6, 0x800025A0}\\
    \texttt{...}     & \texttt{...}       & \texttt{...}       & \texttt{...}           & \texttt{...}\\
    \texttt{M}   & \texttt{45322}     & \texttt{45348}     & \texttt{FBAC5CE3}      & \texttt{BGE S8, S10, 0x800025C0}\\
    \end{tabular}
    }
    \vspace{-0.75cm}
\end{table}

In \textbf{Step 2}, the discrepancies between the snapshots corresponding to the start and end of each speculative window are computed. These discrepancies represent potential information leakage locations, shown in orange in \autoref{fig:framework}. Note that while the microarchitectural state changes due to the execution of the speculative window, not all of these changes indicate the existence of information leakage. Therefore, potential information leakage locations are forwarded to the \vd for detecting direct-channel leakage.

\textbf{\vd} detects direct-channel leakage, i.e., changes in the architectural state due to the execution of a misspeculated window. 
By cross-referencing the altered architectural registers with the PDLC list, we can determine the microarchitectural registers responsible for leaking information to these architectural registers, thereby pinpointing the root cause of the vulnerability. This approach effectively eliminates the need for a golden model, as we leverage the difference snapshot to identify mismatches due to speculative execution. 
If the architectural state has not changed during the speculative window, this only implies that the corresponding input caused no direct leakage from the microarchitectural to the architectural layer. 
However, it does not negate the existence of side-channel leakage, which is out of the scope of this work. \vd provides feedback on the presence of vulnerabilities to guide \hf generating new inputs. 

\textbf{\coc} receives the snapshots and the PDLC list (\autoref{sec:phaseone}) to compute our novel \textit{Leakage Path} (LP) coverage. 
The \textit{LP} metric aims to guide \hf to further explore potential direct leakage channels during speculative execution, thereby increasing the chances of triggering speculative execution vulnerabilities.
It computes the \textit{LP} coverage based on the number of times the PLDC signals toggled during the speculative window.

\section{Evaluation Results}
\label{sec:evaluation}
We evaluated our approach using the most intricate, open-source  RISC-V out-of-order processor, \boom{}~\cite{boom}.
We used Chipyard~\cite{chipyard} as the simulation environment for \boom{}. 
We conducted experiments on a 32-core, 2.6 GHz Intel Xeon processor with 512 GB RAM running Linux-based Cent OS.

\subsection{\textsc{Offline Phase Evaluation}}

\textbf{IFG.} We employed Pyverilog~\cite{takamaeda2015pyverilog} to parse \boom{}'s Verilog code into an abstract syntax tree~(AST) and generate its IFG. 
This step took around 9 minutes for \boom{} and is required once per PUT.
The resulting $IFG (R,F)$ includes $162,631$ signals in $R$ and $428,245$ connections in $F$. 

\noindent \textbf{PDLC.} As outlined in \autoref{sec:phaseone}, we extracted \boom{}'s architectural registers using the RISC-V ISA documentation. 
Subsequently, we employed the Depth-First Search (DFS) algorithm to determine all paths from \boom{}'s microarchitectural to architectural registers in around 3 minutes. The total number of potential direct leakage channels is $9,048$. 

\subsection{\phasetwo~Evaluation}
\coolname effectively detected two of the most recent direct speculative execution vulnerabilities~\cite{wait_for_it,zenbleed}, which cannot be discovered by state-of-the-art hardware fuzzers~\cite{thehuzz,rfuzz,specdoctor,hide_and_seek_with_spectres} because of two main advantages of \coolname: 
(1) a novel vulnerability detection mechanism that employs fine-grained microarchitectural snapshots of the PUT to trace dynamic information flows and spot misbehaviors, i.e., information leakage, due to speculative execution. Owing to this feature, \coolname{} is more accurate than existing fuzzers that rely on comparing execution traces with a golden-reference model~\cite{thehuzz,rfuzz,hide_and_seek_with_spectres} or differential fuzzing \cite{specdoctor}, 
(2) a novel fine-grained coverage metric, \textit{Leakage Path} (LP), which guides the fuzzer toward potential information leakage channels utilizing static IFT, in contrast to general code coverage feedback~\cite{thehuzz,rfuzz,hide_and_seek_with_spectres,specdoctor}.

Since \texttt{(M)WAIT}~\cite{wait_for_it} and \texttt{Zenbleed}~\cite{zenbleed} attacks exploit advanced optimization features, sleep on monitor address and zeroing registers optimization, which are not yet implemented in RISC-V ISA, we first discuss how we emulated their behaviors in \boom{}.
Then, we compare the effectiveness of \coolname with state-of-the-art hardware fuzzers in vulnerability detection \cite{specdoctor,thehuzz}.
Unlike the requirements imposed by \textit{SpecDoctor}~\cite{specdoctor}, \coolname requires no PUT instrumentation or hardware modifications, thereby introducing no additional simulation overhead.
However, \coolname still incurs a runtime overhead of $82\%$ higher than \textit{TheHuzz} \cite{thehuzz} due to snapshots processing and coverage metric computation.
Finally, we demonstrate the efficacy of our coverage metric.

\textbf{Emulating \texttt{(M)WAIT}~\cite{wait_for_it}.} %
We introduced a minor modification
to \boom{} to emulate the \texttt{(M)WAIT} vulnerability. 
The original (M)WAIT attack \cite{wait_for_it} requires the \texttt{umonitor} instruction to invoke a monitor on a memory address, and the \texttt{umwait} instruction sets the core to sleep.
The core wakes up when changes occur at the pertinent memory location or after a predetermined amount of time. 
To maintain generality and avoid unnecessary complexities, we extend the functionality of the RISC-V ISA without introducing new non-standard instructions.
We opted to incorporate three new CSR registers that provide the necessary functionality for emulating the \texttt{(M)WAIT} vulnerability. The new CSRs are \texttt{mwait\_en}, \texttt{monitor\_addr}, and \texttt{mwait\_timer}. 
To emulate \texttt{(M)WAIT} vulnerability behavior on a single-core \boom{}, the user sets the \texttt{monitor\_addr} for monitoring. Activating the waiting behavior involves initiating the timer by writing one to \texttt{mwait\_en}. If changes occur in the monitored memory (\texttt{monitor\_addr}), the \texttt{mwait\_timer} is set to zero. If the timer reaches zero, it is set to one. We modified \boom{}'s data cache to turn off the timer not only with memory location but also with corresponding cache line changes to implement the root cause of the attack.

\textbf{Emulating \texttt{Zenbleed}~\cite{zenbleed}.} To emulate \texttt{Zenbleed}~\cite{zenbleed} in a single-core environment, we simplified the vulnerability constraints while maintaining its generality and root cause, which is the change of a general-purpose register within a mispredicted speculative window that remains after closing the window. To avoid adding non-standard instructions, we introduced a new CSR register, \texttt{zenbleed\_en}. We modified the \texttt{Rename Stage} of the \boom{} pipeline by manipulating the \texttt{maptable} rollback mechanism to prevent the rollback of \texttt{Register File} changes when \texttt{zenbleed\_en} is set to a non-zero value.

Note that emulating the \texttt{(M)WAIT}~\cite{wait_for_it} and \texttt{Zenbleed}~\cite{zenbleed} attacks on a single-core open-source RISC-V processor, inherently lacking advanced optimization features like register zeroing and sleep on monitor address, is a proof of concept for \coolname's efficacy of identifying direct speculative execution vulnerabilities. Our emulation did not dismiss the original vulnerabilities' generality and root cause, i.e., architecture-visible changes during speculative execution.
Moreover, following modifications, we built and tested \boom{} to confirm its functionality correctness.

\begin{table}[t]
\centering
\caption{\coolname{}'s vulnerability detection effectiveness compared to prior works (\textit{``e.m.'' indicates emulated}).}
\vspace{-0.2cm}
\resizebox{0.9\columnwidth}{!}{%
\begin{tabular}{c|c|c|c|c}
Paper                                                        & \begin{tabular}[c]{@{}c@{}}Spectre\\ v1 \end{tabular} & \begin{tabular}[c]{@{}c@{}}Spectre\\ v2\end{tabular} & \begin{tabular}[c]{@{}c@{}}\texttt{(M)WAIT}\\ e.m.\end{tabular} & \begin{tabular}[c]{@{}c@{}}Zenbleed\\ e.m.\end{tabular} \\ \hline

\begin{tabular}[c]{@{}c@{}} \cite{specdoctor} \end{tabular} & \tikzcmark{} & \tikzcmark{} & \tikzxmark{} & \tikzxmark{} \\ %
\begin{tabular}[c]{@{}c@{}} \cite{fadiheh2022exhaustive}\end{tabular} & \tikzcmark{} & \tikzcmark{} & \tikzxmark{} & \tikzxmark{} \\ %
\coolname{} & \tikzcmark{} &  \tikzcmark{} & \tikzcmark{} & \tikzcmark{} \\ %
\end{tabular}
}
\vspace{-0.6cm}
\end{table}

\textbf{Detecting Emulated (M)WAIT and Zenbleed.}
To demonstrate the effectiveness of \coolname, we compared it with \textit{SpecDoctor}~\cite{specdoctor}, the state-of-the-art hardware pre-silicon fuzzer for detecting speculative execution vulnerabilities. We initialized 
both fuzzers in the same condition to run for 24 hours.
After approximately 14 hours and 4.5 hours, \coolname triggered the emulated \texttt{(M)WAIT} and \texttt{Zenbleed} vulnerabilities, respectively. \coolname detected the direct information leakage path as the root cause. The root cause report showed a direct leakage path between the data cache and \texttt{mwait\_timer} CSR register for \texttt{(M)WAIT} and a direct leakage path between \texttt{zenbleed\_en} CSR register and general purpose register file and rename module for \texttt{Zenbleed}. These two vulnerabilities are categorized as CWE-1342.
SpecDoctor \cite{specdoctor} practically could not detect these vulnerabilities within 24 hours. However, considering its vulnerability detection mechanism, which relies on differential fuzzing with varied secrets
and monitors the hash values of instrumented components for mismatches, it cannot detect these vulnerabilities because: (1) limitation in instrumented microarchitectural components, (2) generating random inputs without using a fine-grained coverage metric, and (3) varied secret-based vulnerability detection will overlook information leakages that do not directly reflect the secret value. 

\textbf{Detecting Spectre Vulnerabilities.} We evaluate \coolname's capability in detecting known speculative execution vulnerabilities, i.e., Spectre variant 1 and 2 \cite{spectre}. For this test, we added a data cache to the PDLC list to be monitored by \vd. To assess \coolname's performance, we compared it again against \textit{SpecDoctor} \cite{specdoctor}, which reported that it found Spectre vulnerability in 31 hours. However, \coolname detected this vulnerability $20 \times$ faster than state-of-the-art, after 1.5 hours and 49 minutes, without and with initial seeds with speculative window, respectively.

\begin{figure}[tb]
    \centering
    \includegraphics[width=0.8\linewidth]{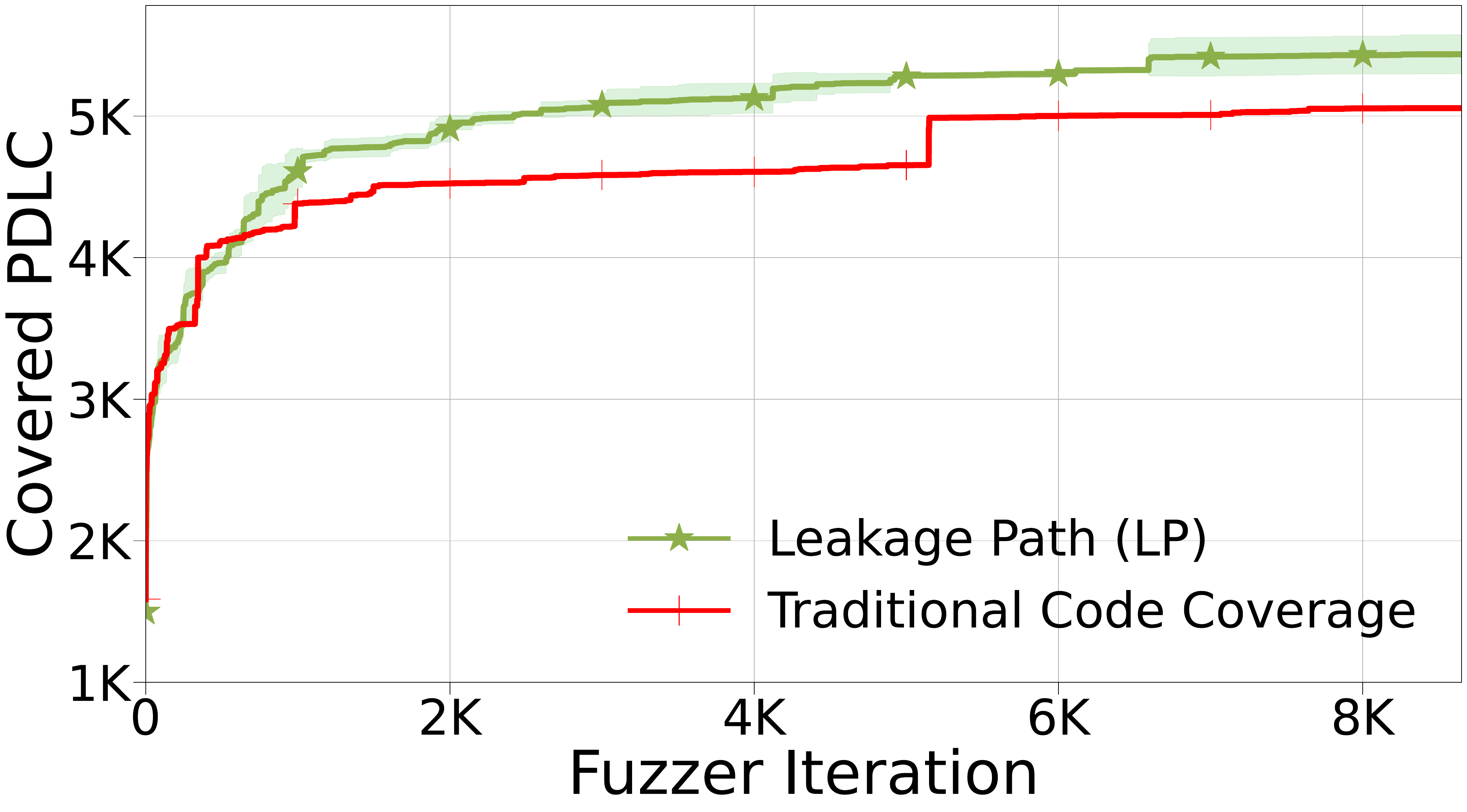}
    \caption{Coverage analysis of traditional code coverage, and \coolname's novel coverage metric.}
    \label{fig:lpc_result}
    \vspace{-0.9cm}
\end{figure}
\textbf{Coverage Analysis.}
We performed a comparative evaluation of novel \textit{LP} coverage against traditional coverage metrics~\cite{thehuzz} in terms of vulnerability detection efficiency, i.e., PDLC activation. In this experiment, we ran \coolname with two different coverage metrics as feedback: (1) novel \textit{Leakage Path}~(LP) and (2) traditional \textit{Code Coverage}~(e.g. FSM, toggle, branch, and condition).
\autoref{fig:lpc_result} presents \boom{}'s covered PDLCs during fuzzing using traditional \textit{code coverage} and \textit{LP} coverage metric. Each experiment was repeated three times, and the plot represents the mean value of all experiments. 
\autoref{fig:lpc_result} shows that the number of explored PDLCs achieved by fuzzer guided by code coverage lags in worst cases by 10.2\% behind the fuzzer guided by \textit{LP} coverage.
Moreover, fuzzer guided by \textit{LP} coverage achieves the same coverage value with only $798$ iterations, compared to the $5149$ iterations required by code coverage guided fuzzer.
This indicates that by employing \textit{LP}, \coolname explores the search space for speculative execution vulnerabilities $6.45\times$  faster than traditional code coverage metrics.

\section{Conclusion}
In this paper, we proposed \coolname, a pre-silicon verification approach that combines hardware fuzzing and IFT to detect speculative execution vulnerabilities in processors. \coolname provided a comprehensive approach for addressing the challenges of integerating hardware fuzzing and IFT. By leveraging the  novel \textit{Leakage Path} coverage metric, \coolname offers an efficient hardware-agnostic solution for speculative information leakage detection. %
In doing so, \coolname contributes to advancing proactive hardware security verification.\blfootnote{\textbf{Acknowledgment.} This work was partially funded by Intel's Scalable Assurance Program, Deutsche Forschungsgemeinschaft (DFG) – SFB 1119 – 236615297, the European Union under Horizon Europe Programme – Grant Agreement 101070537 – CrossCon, the European Research Council under the ERC Programme - Grant 101055025 - HYDRANOS, the US Office of Naval Research (\#N00014-18-1-2058), and the Lockheed Martin Corporation.}

\bibliographystyle{ACM-Reference-Format}
\bibliography{sample-base}


\begin{thebibliography}{20}


\ifx \showCODEN    \undefined \def \showCODEN     #1{\unskip}     \fi
\ifx \showDOI      \undefined \def \showDOI       #1{#1}\fi
\ifx \showISBNx    \undefined \def \showISBNx     #1{\unskip}     \fi
\ifx \showISBNxiii \undefined \def \showISBNxiii  #1{\unskip}     \fi
\ifx \showISSN     \undefined \def \showISSN      #1{\unskip}     \fi
\ifx \showLCCN     \undefined \def \showLCCN      #1{\unskip}     \fi
\ifx \shownote     \undefined \def \shownote      #1{#1}          \fi
\ifx \showarticletitle \undefined \def \showarticletitle #1{#1}   \fi
\ifx \showURL      \undefined \def \showURL       {\relax}        \fi
\providecommand\bibfield[2]{#2}
\providecommand\bibinfo[2]{#2}
\providecommand\natexlab[1]{#1}
\providecommand\showeprint[2][]{arXiv:#2}

\bibitem[et~al.(2021)]%
        {ift_survay}
\bibfield{author}{\bibinfo{person}{H.~Wei et al.}} \bibinfo{year}{2021}\natexlab{}.
\newblock \showarticletitle{Hardware Information Flow Tracking}.
\newblock \bibinfo{journal}{\emph{Comput. Surveys}}.
\newblock


\bibitem[Moghimi(2023)]%
        {moghimi2023downfall}
\bibfield{author}{\bibinfo{person}{Daniel Moghimi}.} \bibinfo{year}{2023}\natexlab{}.
\newblock \showarticletitle{{Downfall}: Exploiting Speculative Data Gathering}. In \bibinfo{booktitle}{\emph{{USENIX} Security}}.
\newblock


\bibitem[Ormandy(2023)]%
        {zenbleed}
\bibfield{author}{\bibinfo{person}{T. Ormandy}.} \bibinfo{year}{2023}\natexlab{}.
\newblock \bibinfo{title}{Zenbleed}.
\newblock \bibinfo{howpublished}{\url{https://bit.ly/zenbleed}}.
\newblock


\bibitem[RISC-V(2021)]%
        {risc_v_isa_0}
\bibfield{author}{\bibinfo{person}{RISC-V}.} \bibinfo{year}{2021}\natexlab{}.
\newblock \bibinfo{title}{The RISC-V Instruction Set Manual Volume I, II}.
\newblock
\newblock


\bibitem[Takamaeda-Yamazaki(2015)]%
        {takamaeda2015pyverilog}
\bibfield{author}{\bibinfo{person}{S. Takamaeda-Yamazaki}.} \bibinfo{year}{2015}\natexlab{}.
\newblock \showarticletitle{Pyverilog: A python-based hardware design processing toolkit for verilog hdl}. In \bibinfo{booktitle}{\emph{ARC}}. Springer.
\newblock


\bibitem[\textit{et al.}(2020a)]%
        {chipyard}
\bibfield{author}{\bibinfo{person}{A.~Amid \textit{et al.}}} \bibinfo{year}{2020}\natexlab{a}.
\newblock \showarticletitle{Chipyard: Integrated Design, Simulation, and Implementation Framework for Custom SoCs}.
\newblock \bibinfo{journal}{\emph{IEEE Micro}}.
\newblock


\bibitem[\textit{et al.}(2020b)]%
        {afl++}
\bibfield{author}{\bibinfo{person}{A.~Fioraldi \textit{et al.}}} \bibinfo{year}{2020}\natexlab{b}.
\newblock \showarticletitle{AFL++: Combining Incremental Steps of Fuzzing Research}. In \bibinfo{booktitle}{\emph{{USENIX} Workshop on Offensive Technologies}}.
\newblock


\bibitem[\textit{et al.}(2018a)]%
        {cache_attack_tlb}
\bibfield{author}{\bibinfo{person}{B.~Gras \textit{et al.}}} \bibinfo{year}{2018}\natexlab{a}.
\newblock \showarticletitle{{Translation leak-aside buffer: Defeating cache side-channel protections with TLB attacks}}. In \bibinfo{booktitle}{\emph{{USENIX} Security}}.
\newblock


\bibitem[\textit{et al.}(2022a)]%
        {solt_cellift_2022}
\bibfield{author}{\bibinfo{person}{F.~Solt \textit{et al.}}} \bibinfo{year}{2022}\natexlab{a}.
\newblock \showarticletitle{{CellIFT: Leveraging Cells for Scalable and Precise Dynamic Information Flow Tracking in RTL}}.
\newblock \bibinfo{journal}{\emph{{USENIX} Security}}.
\newblock


\bibitem[\textit{et al.}(2019a)]%
        {hardfails}
\bibfield{author}{\bibinfo{person}{G.~Dessouky \textit{et al.}}} \bibinfo{year}{2019}\natexlab{a}.
\newblock \showarticletitle{{HardFails: Insights into Software-Exploitable Hardware Bugs}}.
\newblock \bibinfo{journal}{\emph{{USENIX} Security}}.
\newblock


\bibitem[\textit{et al.}(2022b)]%
        {specdoctor}
\bibfield{author}{\bibinfo{person}{J.~Hur \textit{et al.}}} \bibinfo{year}{2022}\natexlab{b}.
\newblock \showarticletitle{SpecDoctor: Differential Fuzz Testing to Find Transient Execution Vulnerabilities}. In \bibinfo{booktitle}{\emph{{ACM} {SIGSAC} CCS}}. \bibinfo{publisher}{ACM}.
\newblock


\bibitem[\textit{et al.}(2015)]%
        {boom}
\bibfield{author}{\bibinfo{person}{K.~Asanovic \textit{et al.}}} \bibinfo{year}{2015}\natexlab{}.
\newblock \bibinfo{booktitle}{\emph{{The Berkeley Out-of-Order Machine (BOOM): An industry-competitive, synthesizable, parameterized RISC-V processor}}}.
\newblock \bibinfo{type}{{T}echnical {R}eport}.
\newblock


\bibitem[\textit{et al.}(2018b)]%
        {rfuzz}
\bibfield{author}{\bibinfo{person}{K.~Laeufer \textit{et al.}}} \bibinfo{year}{2018}\natexlab{b}.
\newblock \showarticletitle{RFUZZ: coverage-directed fuzz testing of RTL on FPGAs}. In \bibinfo{booktitle}{\emph{IEEE/ACM ICCAD}}.
\newblock


\bibitem[\textit{et al.}(2022c)]%
        {fadiheh2022exhaustive}
\bibfield{author}{\bibinfo{person}{M.~Fadiheh \textit{et al.}}} \bibinfo{year}{2022}\natexlab{c}.
\newblock \showarticletitle{An exhaustive approach to detecting transient execution side channels in RTL designs of processors}.
\newblock \bibinfo{journal}{\emph{IEEE Trans. Comput.}}
\newblock


\bibitem[\textit{et al.}(2021)]%
        {introspectre}
\bibfield{author}{\bibinfo{person}{M.~Ghaniyoun \textit{et al.}}} \bibinfo{year}{2021}\natexlab{}.
\newblock \showarticletitle{INTROSPECTRE: A Pre-Silicon Framework for Discovery and Analysis of Transient Execution Vulnerabilities}.
\newblock \bibinfo{journal}{\emph{{ACM/IEEE} Annual International Symposium on Computer Architecture}}.
\newblock


\bibitem[\textit{et al.}(2023a)]%
        {hide_and_seek_with_spectres}
\bibfield{author}{\bibinfo{person}{O.~Oleksenko \textit{et al.}}} \bibinfo{year}{2023}\natexlab{a}.
\newblock \showarticletitle{Hide and Seek with Spectres: Efficient discovery of speculative information leaks with random testing}.
\newblock \bibinfo{journal}{\emph{{IEEE} Symposium on Security and Privacy}}.
\newblock


\bibitem[\textit{et al.}(2022d)]%
        {aepic}
\bibfield{author}{\bibinfo{person}{P.~Borrello \textit{et al.}}} \bibinfo{year}{2022}\natexlab{d}.
\newblock \showarticletitle{{{\AE}PIC} Leak: Architecturally Leaking Uninitialized Data from the Microarchitecture}. In \bibinfo{booktitle}{\emph{{USENIX} Security}}.
\newblock


\bibitem[\textit{et al.}(2019b)]%
        {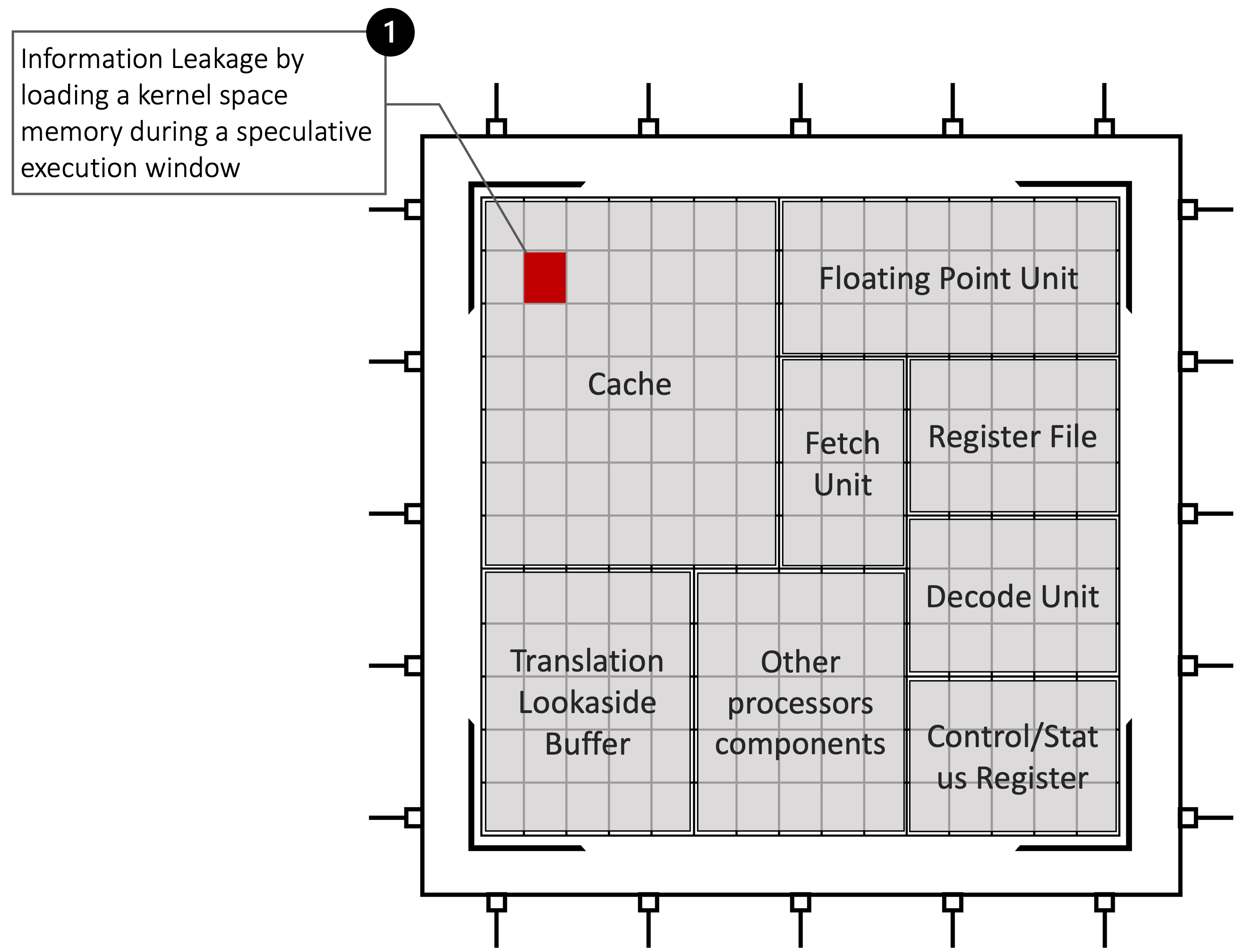}
\bibfield{author}{\bibinfo{person}{P.~Kocher \textit{et al.}}} \bibinfo{year}{2019}\natexlab{b}.
\newblock \showarticletitle{Spectre Attacks: Exploiting Speculative Execution}. In \bibinfo{booktitle}{\emph{{IEEE} Symposium on Security and Privacy}}.
\newblock


\bibitem[\textit{et al.}(2022e)]%
        {thehuzz}
\bibfield{author}{\bibinfo{person}{R.~Kande \textit{et al.}}} \bibinfo{year}{2022}\natexlab{e}.
\newblock \showarticletitle{{TheHuzz}: Instruction Fuzzing of Processors Using {Golden-Reference} Models for Finding {Software-Exploitable} Vulnerabilities}.
\newblock \bibinfo{journal}{\emph{{USENIX} Security}}.
\newblock


\bibitem[\textit{et al.}(2023b)]%
        {wait_for_it}
\bibfield{author}{\bibinfo{person}{R.~Zhang \textit{et al.}}} \bibinfo{year}{2023}\natexlab{b}.
\newblock \showarticletitle{(M)WAIT for It: Bridging the Gap between Microarchitectural and Architectural Side Channels}. In \bibinfo{booktitle}{\emph{{USENIX} Security}}.
\newblock


\end{thebibliography}

\end{document}